\documentclass[]{pasj01}

\Received{2021 April 23}
\Accepted{2021 August 3}
 
\usepackage{graphicx}   
\usepackage{bm}         
\usepackage{natbib}     
\usepackage{aas_macros} 

\newcommand{\minit}{M_{\rm{init}}}
\newcommand{\msun}{\thinspace M_{\odot}}
\newcommand{\gcm}{\thinspace \rm{g} \thinspace \rm{cm}^{-3}}
\newcommand{\gcmcm}{\thinspace \rm{g} \thinspace \rm{cm}^{-2}} 

\newcommand{\kms}{\thinspace \rm{km} \thinspace \rm{s}^{-1}}

\newcommand{\podp}{\psi_{\rm{d1-12}}} 
\newcommand{\pods}{\psi_{\rm{d2-12}}} 
\newcommand{\podt}{\psi_{\rm{d3-12}}} 
\newcommand{\post}{\psi_{\rm{ps3-12}}} 

\begin{document}

\title{
  \LETTERLABEL 
  A new formation scenario of a counter-rotating circumstellar disk:
  spiral-arm accretion from a circumbinary disk in a triple protostar
  system
}

\author{
  Daisuke \textsc{Takaishi}, \altaffilmark{1,}$^{*}$
  Yusuke \textsc{Tsukamoto}, \altaffilmark{1,}$^{*}$
  and
  Yasushi \textsc{Suto} \altaffilmark{2,3,}$^{*}$
}
\altaffiltext{1}{Graduate School of Science and Engineering,
  Kagoshima University, Kagoshima 890-0065, Japan}
\altaffiltext{2}{Department of Physics, The University of Tokyo,
  Tokyo 113-0033, Japan}
\altaffiltext{3}{Research Center for the Early Universe,
  School of Science, The University of Tokyo, Tokyo 113-0033, Japan}
\email{
  k3790238@kadai.jp,
  tsukamoto.yusuke@sci.kagoshima-u.ac.jp,
  suto@phys.s.u-tokyo.ac.jp
}

\KeyWords{
  protoplanetary disks ---
  hydrodynamics ---
  methods: numerical
}

\maketitle

\begin{abstract}
  We present the evolution of rotational directions of circumstellar
  disks in a triple protostar system simulated from a turbulent
  molecular cloud core with no magnetic field.  We find a new
  formation pathway of a counter-rotating circumstellar disk in such
  triple systems.  The tertiary protostar forms via the circumbinary
  disk fragmentation and the initial rotational directions of all the
  three circumstellar disks are almost parallel to that of the orbital
  motion of the binary system.  Their mutual gravito-hydrodynamical
  interaction for the subsequent $\sim10^4\thinspace\rm{yr}$ greatly
  disturbs the orbit of the tertiary, and the rotational directions of
  the tertiary disk and star are reversed due to the spiral-arm
  accretion of the circumbinary disk.  The counter-rotation of the
  tertiary circumstellar disk continues to the end of the simulation
  ($\sim6.4\times10^4\thinspace\rm{yr}$ after its formation), implying
  that the counter-rotating disk is long-lived.  This new formation
  pathway during the disk evolution in Class 0/I Young Stellar Objects
  possibly explains the counter-rotating disks recently discovered by
  ALMA.
\end{abstract}

\section{Introduction}
\label{sec:intro}

Unprecedented recent high angular resolution and sensitivity of the
Atacama Large Millimeter/submillimeter Array (ALMA) have revealed many
surprising discoveries.  Counter-rotating circumstellar disks in young
close binary systems are such examples.  For instance, the O-type
proto-binary system IRAS 16547–4247, of which a projected separation
is $300\thinspace\rm{au}$, hosts counter-rotating twin disks
\citep{2020ApJ...900L...2T}.  Young binary and multiple systems also
frequently have circumstellar disks of which the rotational directions
are misaligned with each other, with that of the
circumbinary/circum-multiple disk, and with the orbital plane
\citep[e.g.,][]{2014Natur.511..567J, 2016ApJ...830L..16B,
  2017ApJ...837...86T, 2019Sci...366...90A, 2020Sci...369.1233K,2021arXiv210611924I}.

While the counter-rotating disks are discovered, the origin has not
been completely clarified from the point of view of the formation
process.  Previous theoretical studies have suggested two major
scenarios of the binary and multiple formation: the turbulent
fragmentation and the disk fragmentation.  The turbulent fragmentation
naturally explains the counter-rotating and/or misaligned disks at the
formation epoch \citep[e.g.,][]{2004ApJ...600..769F,
  2010ApJ...725.1485O}.  This possible formation process of the
primordial counter-rotating and/or misaligned disks, however, prefer
the large separation of $>500\thinspace\rm{au}$
\citep{2010ApJ...725.1485O}.  The IRAS 16547–4247 has the apparent
separation of $300\thinspace\rm{au}$, and the disk fragmentation
scenario is more preferable \citep[e.g.,][]{1989ApJ...347..959A,
  1994MNRAS.271..999B}.  However, it has been believed that the
counter-rotating disks are difficult to form by the disk fragmentation
so far because it tends to form the initially well-aligned disks.

This letter reports a new scenario for the counter-rotating disk
formation in a triple protostar system that forms by the disk
fragmentation; a tertiary circumstellar disk evolves into a
counter-rotating disk by the spiral-arm accretion of the circumbinary
disk.


\section{Outline of our SPH simulation}
\label{sec:Method_and_IC}

We consider a triple protostar system, model C3, listed in Table 1 of
our previous paper (\citet{2020MNRAS.492.5641T}, hereafter Paper I).
Paper I describes details of the numerical simulation.  Therefore, we
briefly present the numerical simulation here.

We perform a set of systematic simulation runs of the isolated
turbulent molecular cloud core collapse using three-dimensional
smoothed particle hydrodynamics (SPH) with $N_{\rm{p}}=10^6$
particles.  The magnetic field is not considered in the simulation.
The formation and evolution of a protostar are represented by the sink
particle \citep{1995MNRAS.277..362B}, which is dynamically created
when the density of an SPH particle exceeds the threshold density
$\rho_{\rm{sink}}=4\times10^{-8}\gcm$ (see section 2.1 in Paper I).

The simulation starts from a cloud core approximated by an isothermal
uniform gas sphere with the initial temperature of
$T_{\rm{init}}=10\thinspace{\rm K}$, the mass of $\minit=1\msun$, and
the radius of $R_{\rm{init}}=3.9\times10^3\thinspace\rm{au}$.  The
initial turbulent velocity field corresponds to the mean Mach number
of ${\cal M}=0.67$, and its velocity power spectrum is proportional to
$k^{-4}$ \citep{2000ApJ...543..822B}, where $k$ is the wavenumber.
Paper I focused on the evolution of the spin-orbit angles of isolated
circumstellar disk systems, and model C3 is one of the two triple
systems generated in Paper I (see their section 2.2 and Table 1).


\section{Formation and evolution of a counter-rotating circumstellar disk}
\label{sec:results}

The simulation is conducted from the cloud core collapse to
$t_{\rm{ps1}}=82,000\thinspace\rm{yr}$, where $t_{\rm{ps1}}$ is the
elapsed time after the primary protostar (ps1) formation.  First, ps1
forms at $\sim49,000\thinspace\rm{yr}$ after the beginning of the
cloud core collapse.  Then, the secondary protostar (ps2) forms at
$t_{\rm{ps1}}=940\thinspace\rm{yr}$ by the disk fragmentation.
Finally, the tertiary protostar (ps3) forms at
$t_{\rm{ps1}}=17,700\thinspace\rm{yr}$ because of the spiral-arm
fragmentation in the circumbinary disk by the gravitational
instability.

\begin{figure*}
  \begin{center}
    \includegraphics[clip,width=170mm]{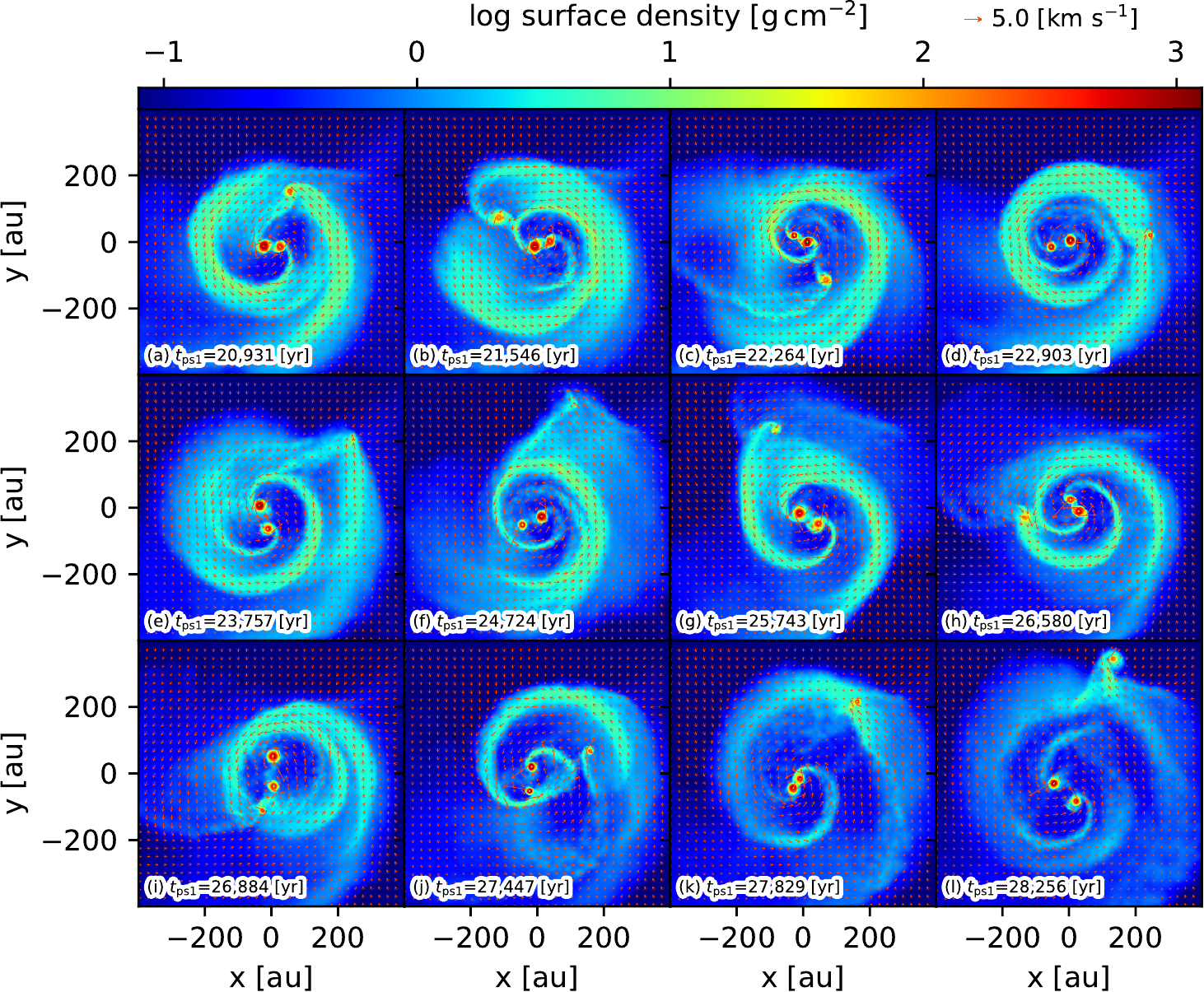}
  \end{center}
  \caption{ Evolution of the surface density along the $z$-direction
    when the rotational direction of the tertiary circumstellar disk
    (csd3) is reversed.  Different panels are labeled by
    $t_{\rm{ps1}}$, the elapsed time after the primary protostar (ps1)
    formation.  The origin of a coordinate system is shifted to the
    center of mass of the triple system.  Red arrows of the
    panels show the density-weighted projected velocity.  The length
    of the arrow is proportional to the velocity amplitude (the
    reference arrow plotted above the figure corresponds to
    $5\thinspace\kms$).
    \label{fig:model_C3_surface_density}
  }
\end{figure*}

\begin{figure}
  \begin{center}
    \includegraphics[clip,width=70mm]{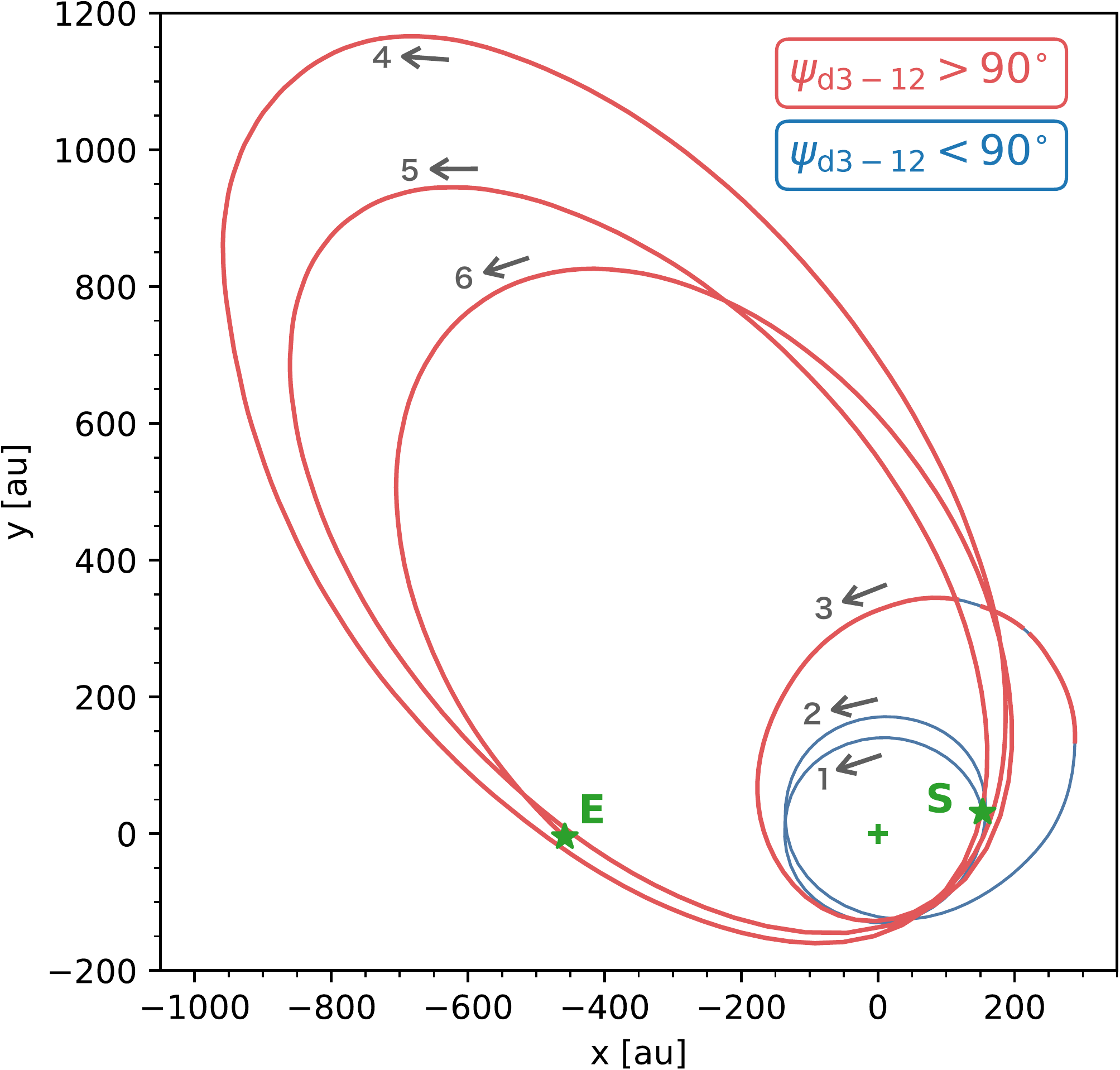}
  \end{center}
  \caption{ Trajectory of the tertiary with respect to the center of
    mass of the inner binary system ($x=y=0$).  The symbols S and E
    mark the positions of csd3 at its formation epoch and at the end
    of the simulation.  Blue and red lines represent the trajectories
    where $\podt<90^{\circ}$ and $\podt>90^{\circ}$, respectively.
    \label{fig:model_C3_orbits}
  }
\end{figure}

Figure \ref{fig:model_C3_surface_density} shows the evolution of the
surface density of the triple system, during which the rotational
direction of the tertiary circumstellar disk (csd3) is reversed in the
evolution.\footnote{ The supplementary movie is available only in the
online edition as ``Supporting Information". } {\bf Panels (a)-(c):}
the rotational direction of csd3 is counter-clockwise and the same as
that of the primary circumstellar disk (csd1), secondary circumstellar
disk (csd2), and the orbital motion of the binary system.  Three
circumstellar disks, csd1, csd2, and csd3, approximately orbit on the
$x$-$y$ plane.  {\bf Panels (c)-(f):} the orbit of csd3 becomes
eccentric, and csd3 starts crossing the circumbinary disk.  Therefore,
the gas component of csd3 is disturbed and partially stripped.  {\bf
  Panels (g)-(j):} csd3 develops by the spiral-arm accretion of the
circumbinary disk, resulting that the rotational direction of csd3 is
reversed (from counter-clockwise to clockwise) and opposite as that of
csd1, csd2, and the orbital motion of the binary system.  {\bf Panels
  (j)-(l):} csd3 passes through the circumbinary disk again and
develops further by the spiral-arm accretion of the circumbinary disk.
Then, csd3 becomes a highly-eccentric and wide-orbit tertiary
companion of the binary system (see also Figure
\ref{fig:model_C3_orbits}).  We find that all the SPH particles of
csd3 in panel (c) of Figure \ref{fig:model_C3_surface_density} are
replenished to newly accreted SPH particles in panel (f) of Figure
\ref{fig:model_C3_surface_density}.  On the other hand, the SPH
particles of csd3 are scarcely replenished in the subsequent
encounters with the spiral arms once the counter-rotation emerges.

\begin{figure*}
  \begin{center}
    \includegraphics[clip,width=155mm]{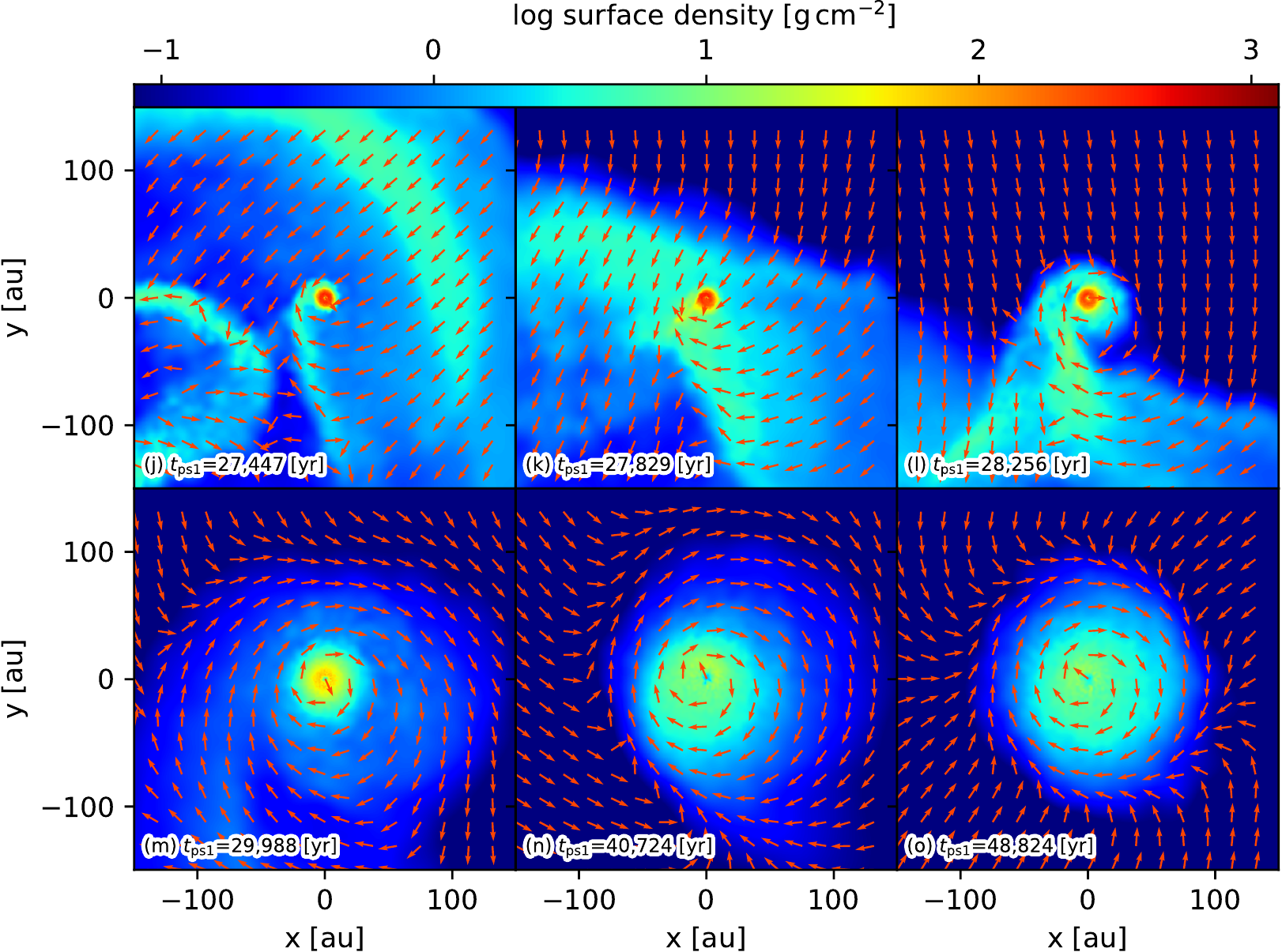}
  \end{center}
  \caption{ Zoom-in view of the evolution of the surface density along
    the $z$-direction.  The origin of a coordinate system is shifted
    to the position of ps3.  Red arrows show the directions of the
    density-weighted projected velocity in the rest frame of ps3.  The
    lengths of the arrows are the same and independent of its velocity
    amplitude.  Panels (j), (k), and (l) correspond to panels (j),
    (k), and (l) in Figure \ref{fig:model_C3_surface_density}.
    \label{fig:model_C3_surface_density_shift_np3}
  }
\end{figure*}

Figure \ref{fig:model_C3_surface_density_shift_np3} shows the zoom-in
view of the surface density around csd3 after the counter-rotation
emerges.  Figure \ref{fig:model_C3_surface_density_shift_np3} clearly
shows that the circumstellar disk is counter-rotating.  Panels (j),
(k), and (l) of Figure \ref{fig:model_C3_surface_density_shift_np3}
indicate that csd3 mainly accumulates the spiral-arm gas that rotates
clockwise around it.  The counter-rotation of csd3 is induced and
enhanced by that spiral-arm gas.  Therefore, we conclude that the
spiral-arm accretion of the circumbinary disk is essential for the
counter-rotating disk formation around the tertiary companion.

\begin{figure}
  \begin{center}
    \includegraphics[clip,width=80mm]{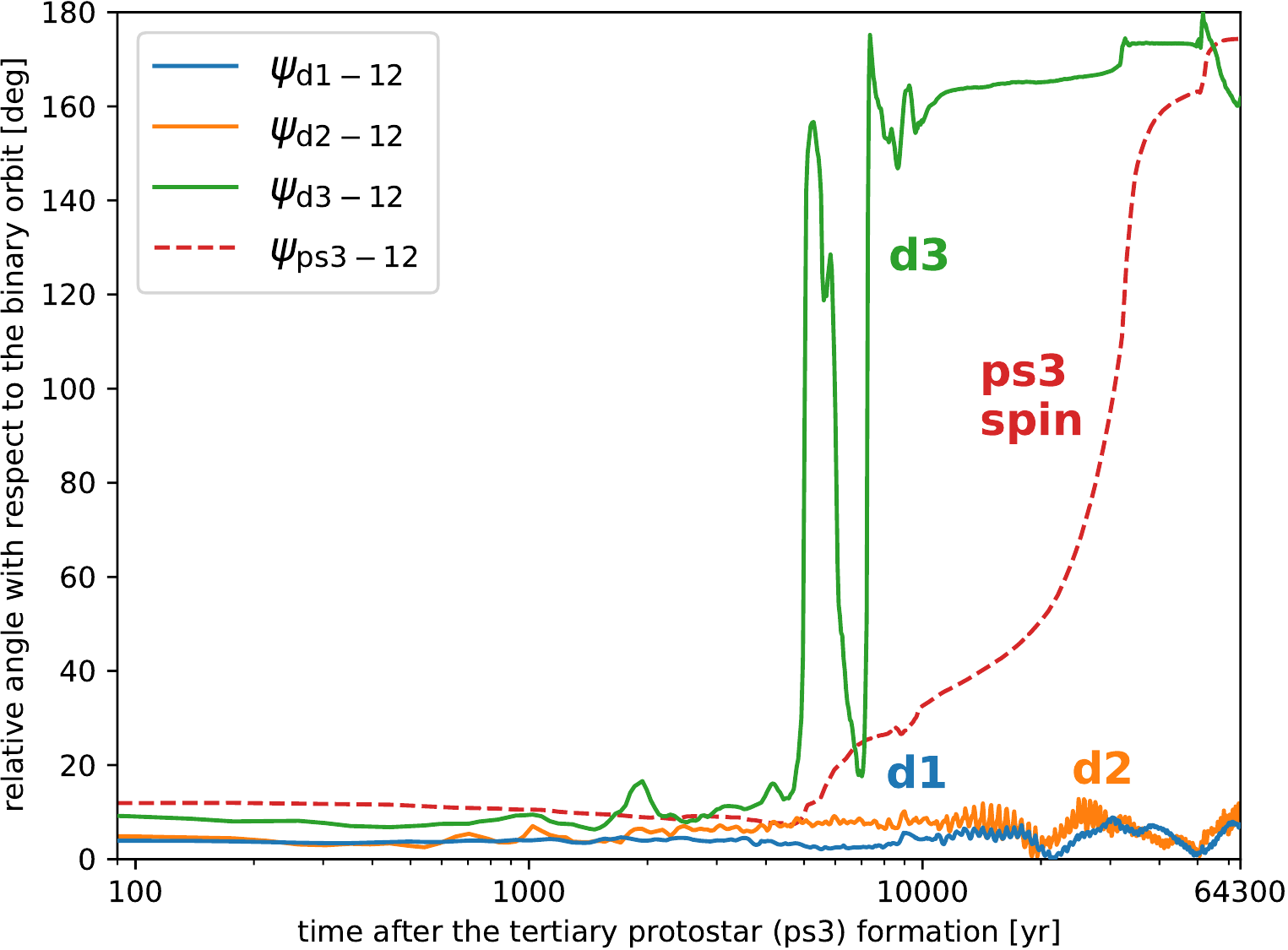}
  \end{center}
  \caption{ Evolution of the disk-orbit angles $\podp$ (blue), $\pods$
    (orange), and $\podt$ (green) as a function of $t_{\rm{ps3}}$ that
    is the elapsed time after ps3 formation.  Red dashed line shows
    the evolution of the star-orbit angle $\post$.
    \label{fig:model_C3_disk_orbit_angle}
  }
\end{figure}

Next, let us consider the evolution of the disk rotational directions
in a more quantitative fashion.  For that purpose, we define the
circumstellar disk as a set of SPH particles that satisfy the
following three conditions: (1) gravitationally bound to the sink
particle, (2) located within $100\thinspace\rm{au}$ from the sink
particle, and (3) their rotational velocity component exceeds the
radial component by a factor of 2
($|(\bm{v}_{\rm{SPH}}-\bm{v}_{\rm{s}})_{t}|>2\thinspace|(\bm{v}_{\rm{SPH}}-\bm{v}_{\rm{s}})_{r}|$,
where $\bm{v}_{\rm{SPH}}$ is the velocity of the SPH particle,
$\bm{v}_{\rm{s}}$ is the velocity of the sink particle, and the
subscripts $t$ and $r$ denote the tangential and radial components of
the relative velocity).  We confirm that our result does not
qualitatively change for other reasonable choices of the disk
definition.

Using the above disk definition, we compute the angular momenta of the
primary, secondary, and tertiary circumstellar disks:
$\bm{J}_{\rm{d1}}$, $\bm{J}_{\rm{d2}}$, and $\bm{J}_{\rm{d3}}$.  Then,
we define the disk-orbit angles relative to the inner binary orbit:
\begin{equation}
  \label{eq:pods}
  \psi_{\rm{d\textit{i}-12}}
  =\cos^{-1}\left(\frac{\bm{J}_{\rm{d\textit{i}}}\cdot\bm{h}_{\rm{12}}}
  {|\bm{J}_{\rm{d\textit{i}}}||\bm{h}_{\rm{12}}|} \right),
  \quad \textit{i}=1,\thinspace2,\thinspace3,
\end{equation}
where, $\bm{h}_{\rm{12}}$ is the specific orbital angular momentum of
the binary system of ps1 and ps2.

Figure \ref{fig:model_C3_disk_orbit_angle} shows the time evolution of
the disk-orbit angles $\podp$ (blue), $\pods$ (orange), and $\podt$
(green) as a function of the elapsed time after the ps3 formation,
$t_{\rm{ps3}}(=t_{\rm{ps1}}-17,700\thinspace\rm{yr})$.  First, the
values of $\podt$ are less than $\sim 20^{\circ}$ in
$t_{\rm{ps3}}\lesssim4\times10^3\thinspace\rm{yr}$.  Then, $\podt$
begins to oscillate in $4\times10^3\thinspace\rm{yr}\leq
t_{\rm{ps3}}\leq10^4\thinspace\rm{yr}$.  Finally, $\podt$ has values
of $\gtrsim 160^{\circ}$ in $t_{\rm{ps3}} \gtrsim
10^4\thinspace\rm{yr}$, meaning that the rotational direction of csd3
is almost antiparallel with that of the orbital motion of the binary
system.  All the values of $\podp$ and $\pods$ are less than
$13^{\circ}$ in $t_{\rm{ps3}}\lesssim 6.4\times
10^4\thinspace\rm{yr}$, meaning that the rotational direction of csd1
and csd2 are both well aligned with that of the orbital motion of the
binary system.  The results clearly show that csd3 is counter-rotating
with respect to csd1, csd2, and the their orbital motion.  The counter
rotation of csd3 continues from $t_{\rm{ps3}}=10^4\thinspace\rm{yr}$
to $t_{\rm{ps3}}=6.4\times10^4\thinspace\rm{yr}$.  Therefore, we are
left with a counter-rotating disk that is stable at least more than
$\sim 5\times10^4\thinspace\rm{yr}$.

Figure \ref{fig:model_C3_disk_orbit_angle} also shows the time
evolution of the star-orbit angle $\post$ that is the relative angle
between the stellar spin axis of ps3 and the orbital axis of the inner
binary system represented by $\bm{h}_{\rm{12}}$.  The stellar spin
axis of ps3 is initially well aligned with the binary orbital axis,
and therefore with the rotation axis of csd3 as well.  Even when csd3
reverses its rotation axis, the stellar spin axis of ps3 remains the
same for a while.  However, the stellar spin axis of ps3 gradually
becomes aligned with the disk rotation axis through the circumstellar
disk accretion to ps3.  Finally, the star-disk angle of the tertiary
system becomes less than $\sim20^{\circ}$, i.e., $\post\approx\podt$.
We confirm that the stellar spin axes of ps1 and ps2 are always well
aligned with their orbital axis within $\sim15^{\circ}$.

Therefore, the present formation scenario of the counter-rotating disk
does not lead to the star-disk misalignment, which can cause the
spin-orbit misalignment often observed for transiting exoplanetary
systems \citep[e.g.,][]{2005ApJ...622.1118O, 2014ApJ...784...66X,
  2015ARA&A..53..409W, 2019AJ....157..137K}.

Figure \ref{fig:model_C3_orbits} plots the trajectory of the tertiary
with respect to the center of mass of the inner binary on the $x$-$y$
plane.  The initially circular orbit becomes eccentric after a few
orbital periods, and its semi-major axis remarkably increases.  The
eccentric orbit of the tertiary plays an essential role in our
spiral-arm accretion scenario for the counter-rotating disk formation.
We find that the triple system satisfies the dynamical instability
condition \citep{2001MNRAS.321..398M} just before the onset of the
eccentric orbit.  Furthermore, the triple system remains unstable and
is still evolving by the hydrodynamical disk interaction and envelope
accretion after the counter-rotation emerges.  Thus, the orbit
instability of the tertiary may be mainly triggered by the
gravitational three-body effect, while the hydrodynamical disk
interaction and envelope accretion are also influential in its
subsequent evolution.

\begin{figure}
  \begin{center}
    \includegraphics[clip,width=0.9\columnwidth]{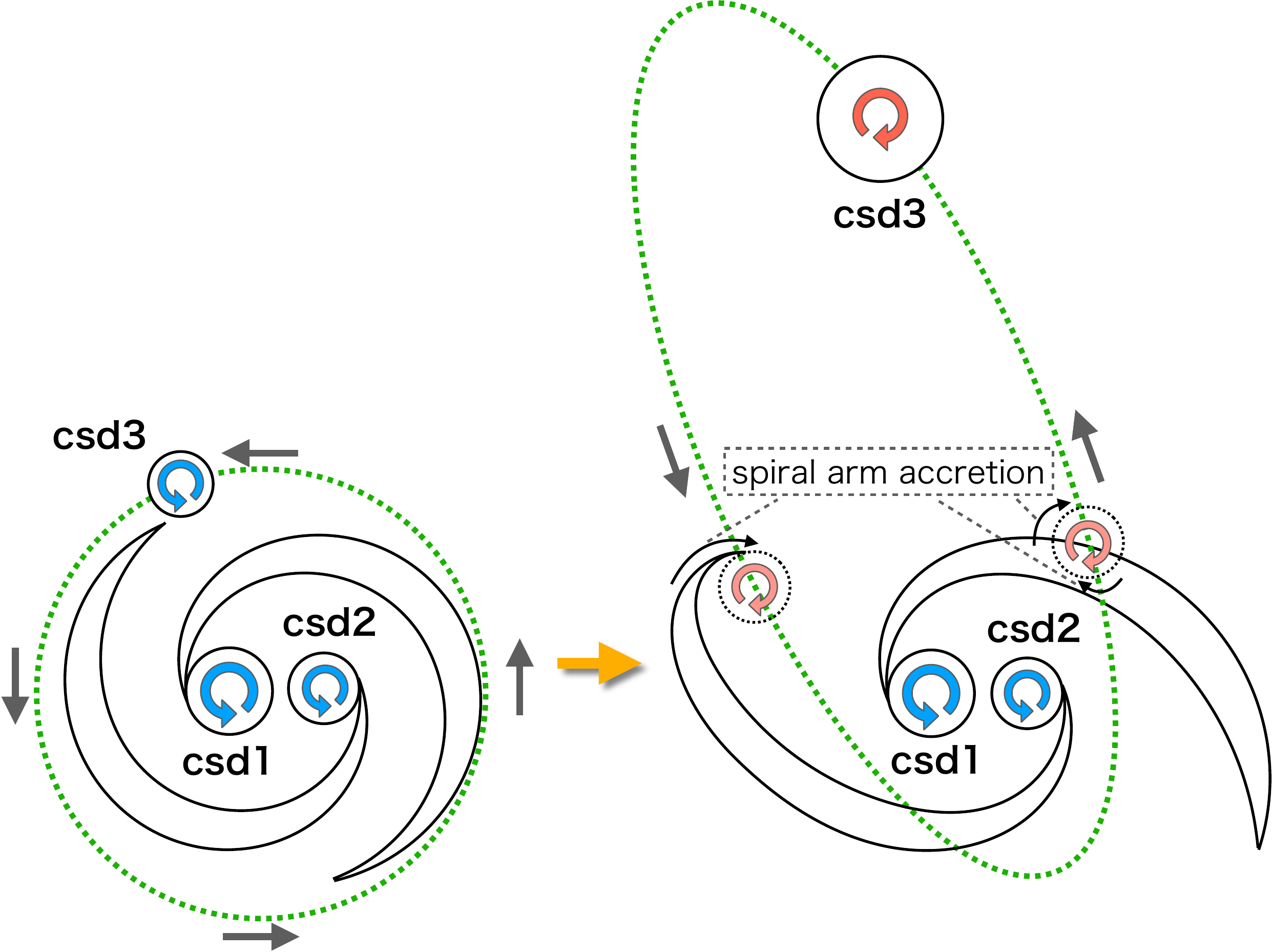}
  \end{center}
  \caption{ Schematic illustration of the spiral-arm accretion
    scenario for the counter-rotating disk formation.
    \label{fig:model_C3_scimatic_figure}
  }
\end{figure}

Figure \ref{fig:model_C3_scimatic_figure} schematically summarizes our
formation scenario of the counter-rotating disk.  The presence of the
circumbinary disk and the eccentric orbital evolution of the tertiary
are the two essential ingredients for the efficient spiral-arm
accretion.

\citet{2018ApJ...869L..44K} discovered a hierarchical triple system HT
Lup: a close ($25\thinspace\rm{au}$) binary system HT Lup A-B with
counter-rotating twin disks has a wide-separation companion HT Lup C
located at $434\thinspace\rm{au}$ away.  Interestingly, the
architecture of HT Lup seems to be similar to the architecture
illustrated in Figure \ref{fig:model_C3_scimatic_figure}.

\begin{figure*}
  \begin{center}
    \includegraphics[clip,width=155mm]{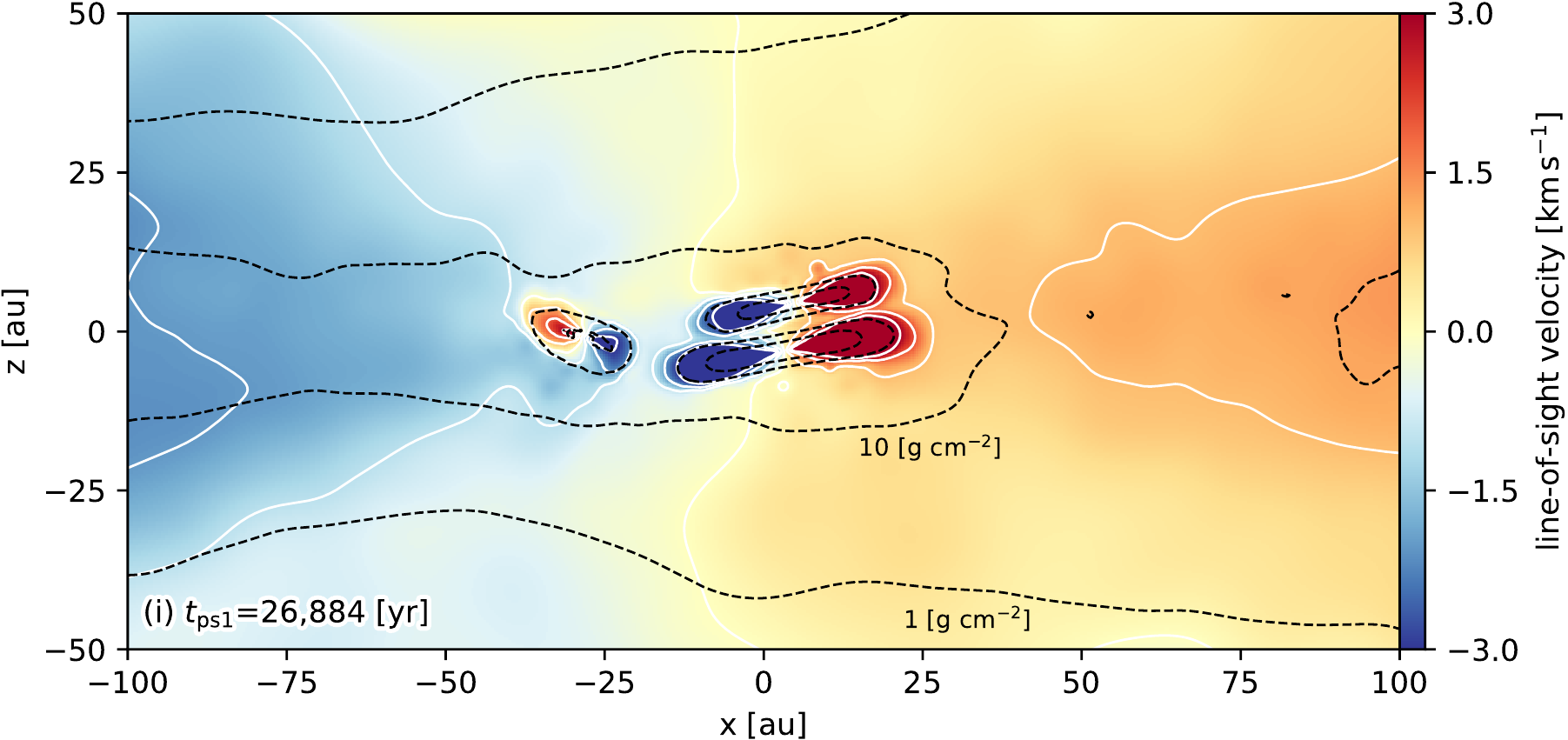}
  \end{center}
  \caption{ Density-weighted line-of-sight velocity viewed from the
    $y$-direction at $t_{\rm{ps1}}=26,884\thinspace\rm{yr}$
    corresponding to panel (i) of Figure
    \ref{fig:model_C3_surface_density}.  The origin of a coordinate
    system is shifted to the center of mass of the triple protostar
    system.  White solid lines represent the contours of the
    density-weighted line-of-sight velocity from $-3\kms$ to $+3\kms$
    in steps of $1\kms$.  Black dashed lines are the contours of the
    surface density at $1\gcmcm$, $10\gcmcm$, $10^{2}\gcmcm$, and
    $10^{3}\gcmcm$ from outer to inner regions.
    The integrations of $y$-direction are taken from
    $-400\thinspace\rm{au}$ to $+400\thinspace\rm{au}$ (i.e., the
    size of the panel (i) of Figure \ref{fig:model_C3_surface_density}).
    \label{fig:model_C3_line_of_sight_vel}
  }
\end{figure*}

Finally, Figure \ref{fig:model_C3_line_of_sight_vel} shows the
density-weighted line-of-sight velocity viewed from the $y$-axis of
Figure \ref{fig:model_C3_surface_density}(i).  The blueshifted and
redshifted velocity structures of the three circumstellar disks appear
to be consistent with the Keplerian rotating disk with the infalling
envelope \citep{1998ApJ...504..314M}.  The velocity pattern of the
blueshifted to redshifted components of csd3 (left) is almost opposite
to that of csd1 (lower right) and csd2 (upper right), indicating that
the projected rotational direction of csd3 is close to antiparallel
with that of csd1 and csd2.  Indeed, this velocity pattern is similar
to observed counter-rotating disks
\citep{2018ApJ...869L..44K, 2020ApJ...900L...2T, 2021arXiv210611924I}.


\section{Conclusion and discussion}
\label{sec:conclusion_and_discussion}

The new formation pathway of the counter-rotating disk (see Figure
\ref{fig:model_C3_scimatic_figure}) is summarized as follows; (i) a
tertiary protostar forms via the fragmentation of the circumbinary
disk, (ii) the tertiary orbit becomes unstable due to the three-body
effect and acquires a large eccentricity, (iii) the circumstellar disk
of the tertiary orbit receives the counter-rotating gas accretion when
crossing the spiral-arm of the circumbinary disk, and its rotational
direction is reversed.  This scenario works on the basis of
gravito-hydrodynamical processes alone and does not require more
complicated physical processes such as the magnetic field.  The
material of the counter-rotating disk is entirely newly accreted, and
it is scarcely replenished in the subsequent encounters with the
spiral arms once the counter-rotating disk forms.

The occurrence rate of the found formation pathway is still unclear so
far.  The counter-rotating disk is formed when the angular momentum of
the accreting fluid elements from the spiral arm to the tertiary is
reversed in the rest frame of the tertiary protostar.  This process
occurs when the tertiary passes through the backside of the spiral
arm.  Thus, it is crucial that whether the
passing-through-the-back-side-of-spiral-arm occurs by chance or with a
relatively high probability to clarify whether the formation of the
counter-rotating disk is rare or not.

The qualitative idea for the counter-rotating disk formation is
speculated as follows.  The tertiary system is formed in the
circumbinary disk, and its orbital period is roughly the same as that
of its natal circumbinary disk even after the tertiary orbit becomes
eccentric.  Thus, the tertiary is expected to be in a 1:1 resonance
with the spiral arms of the circumbinary disk.  Therefore, it is
speculated that the passing-through-the-back-side-of-spiral-arm occurs
in the course of the multiple encounters.  This qualitative idea for
the counter-rotating disk formation will be examined in the follow-up
studies.

Another question is how long the counter-rotating disk survives the
subsequent evolution.  The triple system is still unstable due to the
three-body effect and evolving by the hydrodynamical disk interaction
and envelope accretion after the counter-rotating disk formation.  For
instance, recent theoretical studies suggest that the disk rotational
direction can dynamically change by the envelope accretion that has
the random gas motion by the turbulence \citep{2013MNRAS.428.1321T}
and by the tidal interaction between the binary pair
\citep{2015MNRAS.449L.123M}.  However, we can not currently address
the long-term prospects for the stability and survivability of this
system from the data we have.  The long-term prospects for stability
and survivability will be also given in the follow-up studies.

The remaining questions are crucial to understand if our proposed
scenario can be a major formation channel for a variety of
counter-rotating and misaligned disk systems recently discovered by
ALMA.


\bigskip
\noindent
\textbf{Supporting Information} \\
\noindent
Additional Supporting Information is available in the online version
of this article: The supplementary movie 1.

\begin{ack}
  We are grateful to the anonymous referee for the fruitful
  suggestions and insightful comments that helped to improve the
  manuscript.  Numerical computations were carried out on Cray XC50 at
  Center for Computational Astrophysics, National Astronomical
  Observatory of Japan.  This research is supported by the Japan
  Society for the Promotion of Science (JSPS) Core-to-Core Program
  ``International Network of Planetary Sciences'', and by JSPS KAKENHI
  Grant Numbers JP18H01247 (Y.S.), JP18H05437 (Y.T.), JP18K13581
  (Y.T.), JP19H01947 (Y.S.), and JP21J23102 (D.T.).  The supplementary
  movie was produced using SPLASH \citep{2007PASA...24..159P}.
\end{ack}

\bibliographystyle{apj}  
\bibliography{export-bibtex}  
\end{document}